\documentclass[journal=jacsat,manuscript=article]{achemso}
\usepackage[colorlinks=true,allcolors=blue]{hyperref}
\usepackage{graphicx}
\usepackage{amsmath}
\usepackage{bm}
\usepackage[T1]{fontenc}
\usepackage[latin9]{inputenc}
\usepackage{textcomp}
\usepackage{amstext}
\usepackage[Symbol]{upgreek}
\usepackage{booktabs}
\usepackage{dcolumn}
\usepackage{color}
\usepackage{enumerate}
\usepackage{latexsym,bm}
\usepackage{setspace}
\usepackage{float}
\usepackage{tabularx}

\title{Lattice Defects and the Mechanical Anisotropy of Borophene}

\author{V. Wang}
\affiliation[Xi'an University of Technology]{Department of Applied Physics, Xi'an University of Technology, Xi'an 710054, China}  
\email{wangvei@icloud.com.}
\author{W. T. Geng}
\affiliation{School of Materials Science \& Engineering, University of Science and Technology Beijing, Beijing 100083, China}
\email{geng@ustb.edu.cn.}

\begin{document}

\begin{abstract}
Using density functional theory combined with a semi-empirical van der Waals dispersion correction, we have investigated the stability of lattice defects including boron vacancy, substitutional and interstitial \emph{X} ($X$=H, C, B, N, O) and $\Sigma$5 tilt grain boundaries in borophene and their influence on the anisotropic mechanical properties of this two-dimensional system.
The pristine borophene has significant in-plane Young's moduli and Poisson's ratio anisotropy due to its strong and highly coordinated B-B bonds. The concentration of B vacancy and $\Sigma$5 grain boundaries could be rather high given that their formation energies are as low as 0.10 eV and 0.06 eV/{\AA} respectively. In addition, our results also suggest that borophene can react easily with H$_2$, O$_2$ and N$_2$ when exposed to these molecules. We find that the mechanical strength of borophene are remarkably reduced by these defects. The anisotropy in Poisson's ratio, however, can be tuned by some of them. Furthermore, the adsorbed H or substitutional C may induce negative Poisson's ratio in borophene, and the substitutional C or N can significantly increase the Poisson's ratio by constrast. 
\begin{tocentry} 
\includegraphics[scale=0.28]{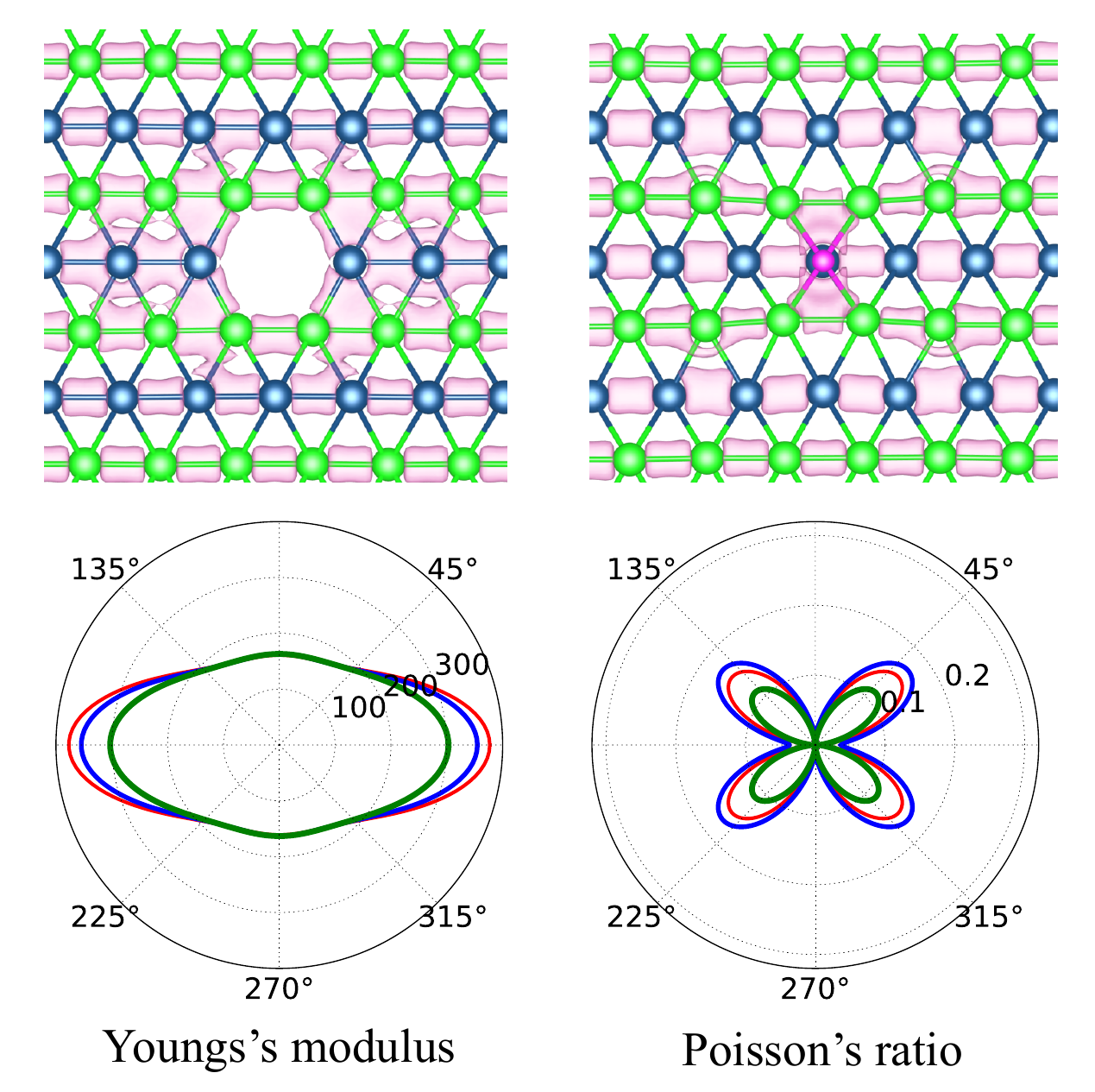}
\end{tocentry}
\end{abstract}

%\linenumbers

\section{Introduction}
Since the successful synthesis of graphene,\cite{Novoselov2004,Novoselov2005,Schwierz2010} two dimensional (2D) materials have attracted tremendous attentions due to their novel electronic, optical, thermal, and mechanical properties for potential applications in various fields.\cite{Song2012,Wang2012c,Zhang2005,Chhowalla2013,Xu2013,Butler2013,Zeng2015,Balendhran2015,Wang2015,Liu2015,Wang2015a,Wang2015b} Boron is the element in the left side of carbon in the periodic table, has three valance electrons and forms in at least 16 allotropes with abundant promising properties. Inspired by the fabrication of graphene, several types of 2D boron nanosheets have been proposed theoretically.\cite{Oganov2009}
Until very recently, a new 2D boron sheet, named as borophene, has been grown successfully
on Ag(111) surfaces under ultrahigh-vacuum conditions and attracted enormous interest due to their extraordinary anisotropic mechanical, electronic, optical and thermal properties.\cite{Mannix2015} For example, both experimental and theoretical studies show that borophene shows metallic conduction along  \emph{x} direction while remains semiconducting along \emph{y} direction (Figure \ref{struct}). Peng \emph{et al.} predicted by performing density functional theory (DFT) calculations that the optical properties of borophene exhibit strong anisotropy. It is therefore a promising candidate for future design of transparent conductors used in photovoltaics with its high optical transparency and high electrical conductivity.\cite{Peng2016} Mannix \emph{et al.} found that the in-plane Young's modulus is 170 GPa$\cdot$nm along \emph{y} direction, but is as high as 398 GPa$\cdot$nm  along \emph{x} direction, exceeding graphene's Young's modulus of 340 GPa$\cdot$nm.\cite{Lee2008,Tsafack2016} With these aspects, borophene is expected to server as important ingredients in future nanodevices.

However, the development of nanoscale electronics has been hindered by the lack of control over the large-scale synthesis of single-crystalline 2D materials. During the material growth, native defects, impurities and extrinsic dopants are inevitably or purposefully introduced. Compared to their bulk counterparts, 2D material defects tend to have more significant influence on their material properties.\cite{Terrones2012,Peng2013,Komsa2014,Lehtinen2003,Jena2007,Liu2007,Lee2008,Han2013,Qiu2013,Ghorbani2013,Cheng2013,Tongay2013,Banhart2011,Liu2014,Wang2015c}
In addition, 2D materials produced so far are typically polycrystalline. Grain boundaries (GBs) play a crucial role in determining mechanical, electrical, optical and magnetic properties, often in connection with performance degradation. For instance, GBs can decrease the electronic and thermal conductance of graphene,\cite{Yazyev2010,Yu2011,Cai2010} enhance or weaken the mechanical strength of graphene.\cite{Grantab2010,Huang2011,Wei2012} Yakobson \emph{et al.} reported theoretical evidence that certain types of GBs introduce midgap electronic states and act as sinks for the carriers in single-layer transition metal disulfides.\cite{Zou2013,Zhou2013} On the other hand, GBs can also be electronically inactive in phosphorene due to the single-element bonding.\cite{Liu2014} The weak screening and reduced dimensionality further magnify the GBs effects in 2D systems.\cite{Grantab2010,Huang2011,Wei2012,Yazyev2010,Yu2011,Liu2010,Kim2011,Liu2012}

To the best of our knowledge, however, little is known about the structure and stability of these defects in borophene. The aim of this paper is to achieve a full understanding of the fundamental features and the influence of lattice defects, including B vacancy, substitutional and interstitial \emph{X} ($X$=H, C, B, N, O) defects, as well as symmetric tilt GBs on the mechanical strength of borophene by performing DFT calculations. 
Our results demonstrate all defects considered in our present study reduce the in-plane Young's moduli of borophene. The adsorption of H or substitution of C for B could cause very negative Poisson's ratio in borophene, and the substitution of C or N for B can significantly increase the Poisson's ratio of the host. 
The remainder of this paper is organized as follows. In Sec. II, the computational method and details are described. Sec. III presents the calculations of structure and stability of point defects in borophene, followed by the discussion of the role of various point defects on the orientation-dependent mechanical properties of borophene. Sec. IV presents GB results using similar strategy as in Sec. III. Finally, a short summary is given in Sec. V.

\section{Computational Details}
Our total energy and electronic structure calculations were performed using the Vienna Ab initio Simulation Package (VASP).\cite{Kresse1996, Kresse1996a} The electron-ion interaction was described using projector augmented wave (PAW) method \cite{PAW, Kresse1999} and the exchange and correlation were treated with generalized gradient approximation (GGA) in the Perdew Burke Ernzerhof (PBE) form\cite{Perdew1996}.
A cutoff energy of 400 eV was adopted for the plane wave basis set, which yields total energies convergence better than 1 meV/atom. 
Considering that the GGA-PBE fails in the description of weakly bound systems such as layered materials where van der Waals (vdW) interactions are the dominant part in the cohesive energy, the vdW interactions were incorporated by employing a semi-empirical correction scheme of Grimme's DFT-D2 method in this study, which has been successful in describing the geometries of various layered materials.\cite{Grimme2006, Bucko2010}

\begin{figure}[H]
\centering
\includegraphics[scale=0.10]{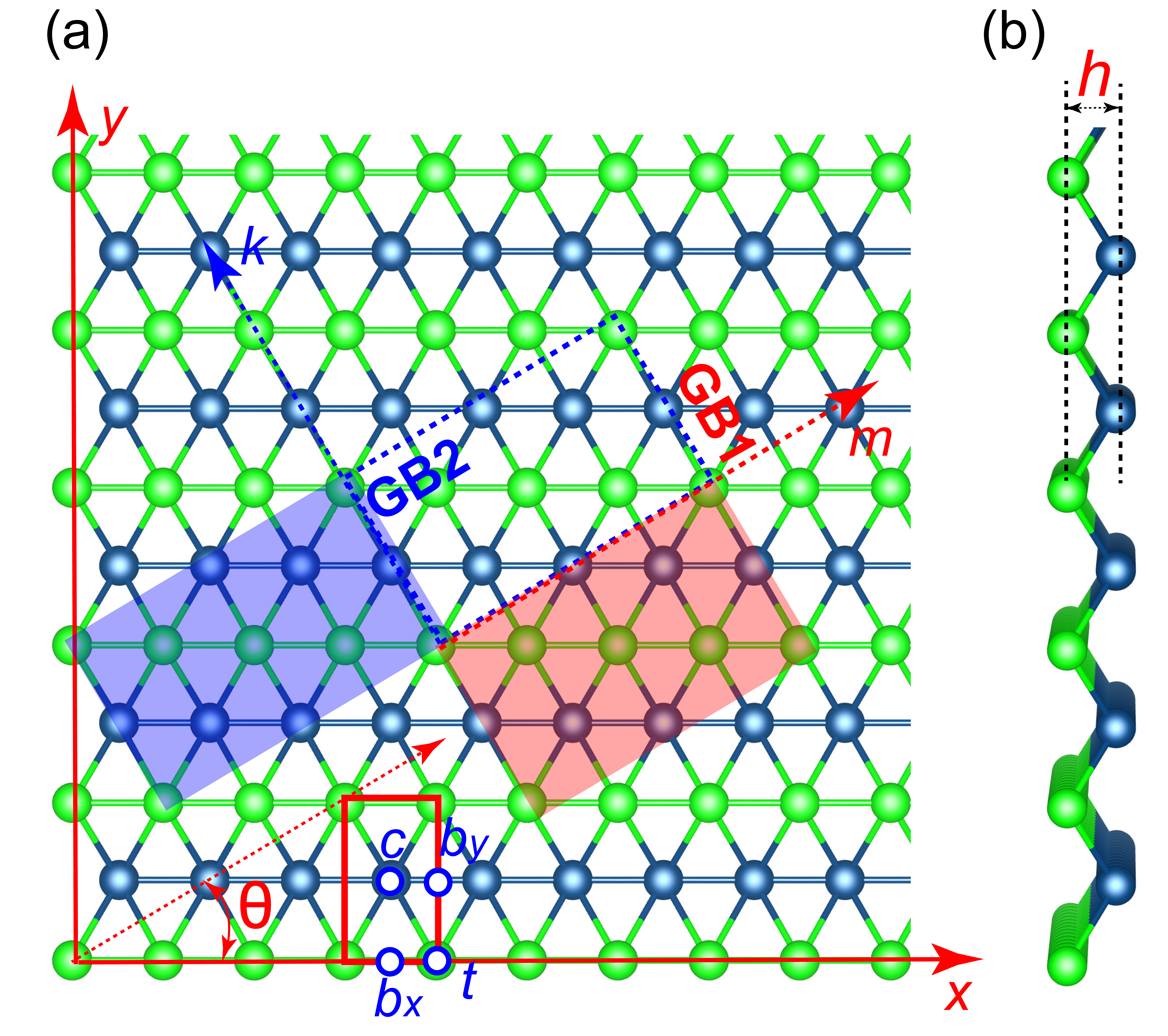}
\caption{\label{struct}(Color online) (a) Top and (b) side views of the unstressed monolayer borophene. The primitive cell is indicated by the red box. Boron atoms in top and in bottom layers are shown as green and blue spheres, respectively.}
\end{figure}

In the slab model of borophene, periodic slabs were separated by a vacuum layer of 15 {\AA} in \emph{c} direction to avoid mirror interactions. In sampling the Brillouin zone integrations, we used Monkhorst-Pack \emph{k}-point meshes with a reciprocal space resolution of 2$\pi$$\times$0.04 {\AA}$^{-1}$.\cite{Monkhorst1976} On geometry optimization, both the shapes and internal structural parameters of pristine unit-cells were fully relaxed until the residual force on each atom is less than 0.01 eV/{\AA}. The atomic structure of the monolayer borophene is presented in Figure \ref{struct}. One can observe that the B atoms form line-chains along \emph{x} direction, while the bonds between B in different chains lead to corrugation along \emph{y} direction. As a result, borophene has a highly anisotropic honeycomb structure with space group \emph{Pmmn}.
For comparison, the B atoms in bottom and top layers are distinguished as green and bule spheres in Figure \ref{struct}, and labeled as B1 and B2 respectively. 
The optimized lattice constants are \emph{a}=1.63 {\AA} and \emph{b}=2.82 {\AA}, in good agreement with both experimental and previous theoretical results.\cite{Mannix2015} The B-B bond length along \emph{x} is 1.63 {\AA}, 0.2 {\AA} smaller than that along \emph{y}
(1.85 {\AA}). It is therefore expected that B-B bond strength along \emph{x} would be stronger that that along \emph{y}. The buckling height \emph{h} is predicted to be 0.87 {\AA}.

To investigate the influence of point defects on the mechanical properties of borophene, a defect is introduced by adsorbing one \emph{X} atom ($X_i$), substituting \emph{X} for B ($X_\text{B}$) or removing a B atom ($V_\text{B}$) from a 5\texttimes{}3 supercell of borophene which consists of 30 atoms, hence a defect concentration of around 3.3\%. Here \emph{X} can be H, C, N and O which are 
often either unintentionally or purposefully incorporated during the growth of semiconductors and other two-dimensional materials.   
For each type of $X_i$ defect, four inequivalent adsorption sites are investigated, namely, the center site of a rectangular unit-cell, the top site directly above a boron atom, and the bridge site above the midpoint of a B-B bond along \emph{x} or \emph{y} direction. We label these sites as \emph{c}, \emph{t}, \emph{b}$_x$ and \emph{b}$_y$ respectively, as shown in Figure \ref{struct}. In view of the fact that the vdW interaction has remarkable contribution to the stability of adsorbate on graphene, even in the chemisorption cases,\cite{Wang2012} we expect that the PBE plus DFT-D2 method should also give a more accurate description on the local structure of interstitial defects in borophene. Both the shapes and internal structural parameters of  the defective supercells were optimized to reduce the residual force on each atom to less than 0.01 eV/{\AA}. Moreover, we have allowed spin-polarization for defective systems to determine their possible magnetic ground states.

\section{Native point defects and impurities}
We begin our investigation by computing the stability of various possible defects in borophene. We have only considered the charge-neutral defects since the charged defects cannot be stable in a metallic system. The temperature- and pressure-dependent formation energy of a neutral defect is defined as
\begin{equation}\label{eq1}
\Delta H_{m}=E_{tot}(X)-E_{tot}(host)-n_{X}\mu_{X}^{T,P},
\end{equation}
where $E_{tot}(X)$ and $E_{tot}(host)$ are the total energies of the supercells with and without defect. \emph{n}$_X$ is the number of atoms of species \emph{X} added to (\emph{n}$_X$>0) or removed from (\emph{n}$_X$<0) the perfect supercell to create defect. $\mu_X$ is the atomic chemical potential of species \emph{X}. 

Under the standard state pressure $\emph{P}_0$=1 Bar, the temperature-dependent chemical potential of $\mu_{X}$ with contributions from enthalpy \emph{H} and entropy \emph{S} is expressed as
\begin{equation}\label{eq2}
\mu_{X}^{T,P_0}=\mu_{X}^{T_0,P_0}+\Delta\mu_X^{T,P_0},
\end{equation}
where $\mu_{X}^{T_0,P_0}$ is the chemical potential of $\mu_{X}$ at the reference temperature $T_0$=298.15 K. It is approximately equal to the total energy of per atom at \emph{T}=0 K, in the most stable elemental phase of \emph{X}, $i.e.$, O$_2$, H$_2$, N$_2$ and graphite for O, H, N and C respectively. $\Delta\mu_X^{T,P_0}$ is the change in chemical potential from $T_0$ to temperature T>$T_0$. We take the case of nitrogen as an example, the $\Delta\mu_{N}^{T,P_0}$ can be calculated by \cite{Osorio2006} 

\begin{equation}\label{eq3}
\begin{aligned}
\Delta\mu_\text{N}^{T,P_{0}}=\frac{1}{2}[H_{0}^{T_0,P_0}+\Delta H^{T,P_{0}}]-\\
\frac{T}{2}[S_{0}^{T_0,P_0}+\Delta S^{T,P_{0}}],
\end{aligned}
\end{equation}
where the enthalpy $H_0$=8.67 kJ$\cdot$mol$^{-1}$ and the entropy $S_0$=191.61 J$\cdot$K$^{-1}$$\cdot$mol$^{-1}$ for N$_2$ atom at the standard state, namely \emph{T}$_0$=298.15 K and \emph{P}$_0$=1 Bar. We also have $\Delta H^{T, P_0}$=$C_P^{T, P_0}(T-T_0)$, $\Delta S^{T, P_0}$=$C_P$ln($T$/$T_0$), $C_P$=3.5 $k_\text{B}$ for the constant-pressure heat capacity per diamolecule. These temperature-dependent parameters can be taken from the thermochemical tables.\cite{Haynes2015} Under other ambient pressures, we have $\mu_\text{N}^{T,P}$=$\mu_\text{N}^{T, P_0}$+$k_\text{B}$/2$\cdot$$T$ln($P$/$P_0$).\cite{Reuter2001} 

\begin{figure}[H]
\centering
\includegraphics[scale=0.60]{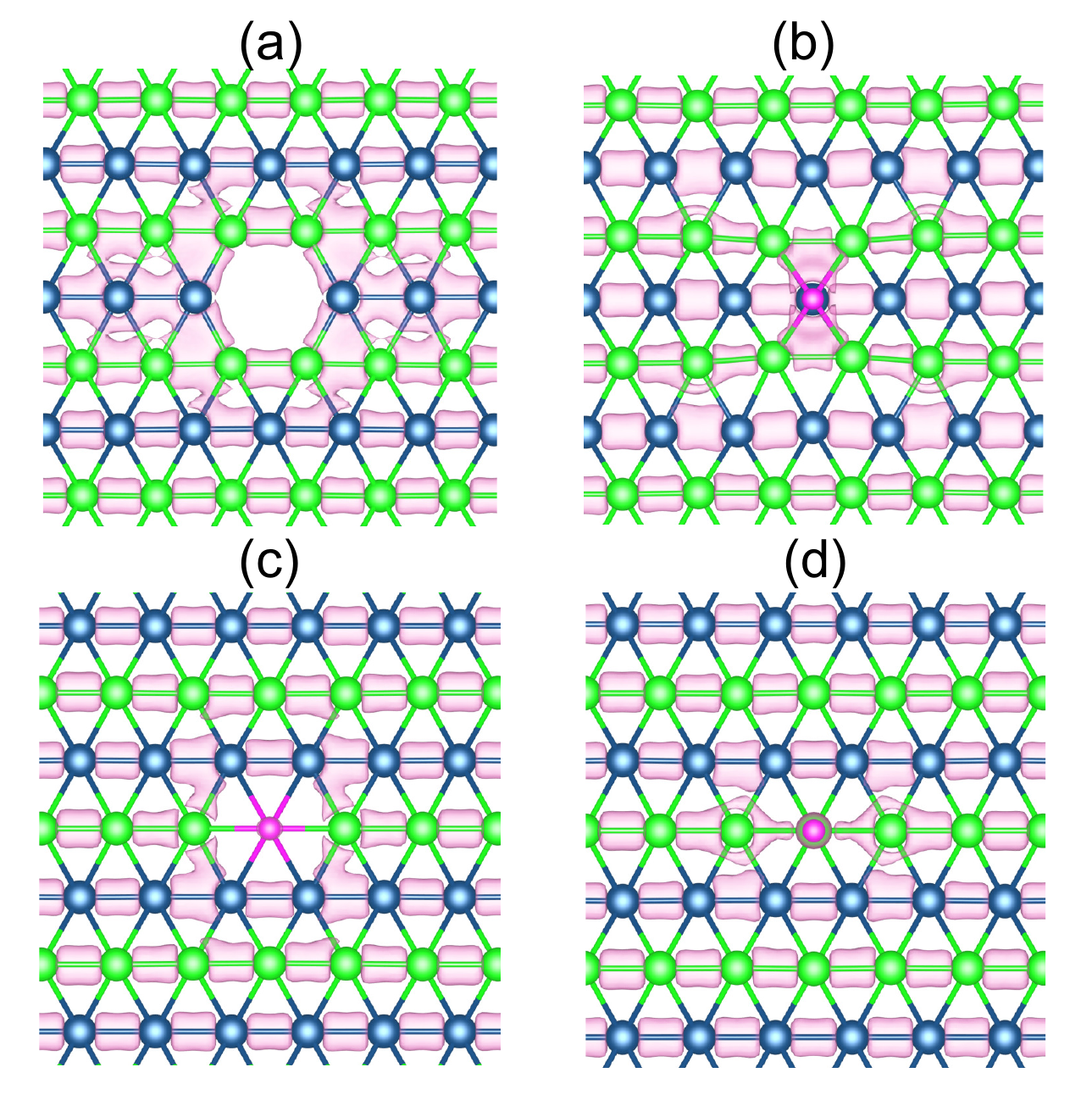}
\caption{\label{chg-B-H}(Color online) Charge density plots of the energetically most stable defective borophene consisting of one (a) V$_\text{B}$, (b) B$_i$, (c) H$_\text{B}$, or (d) H$_i$ defect, respectively. The B1 and B2 are colored with green and blue. The isosurface is 0.12 e/Bohr$^3$. }
\end{figure}

From the calculated results listed in Table \ref{tab1}, one can find that V$_\text{{B}}$ is the preferable native defect with a formation energy of 0.10 eV at the standard state, more stable than that of B$_i$ by 0.90 eV. 
To gain knowledge of the distribution of V$_\text{{B}}$ at a concentration of 3.3\%, we have investigated three kinds of V$_\text{{B}}$ alighments in a 90-atom supercell: (i) uniform distribution, (ii) three V$_\text{{B}}$ clustered to form a trivacancy, and (iii) disordered distribution. 
The disordered V$_\text{{B}}$ distribution is modeled based on the so-called special quasi-random structure (SQS) approach.\cite{Zunger1990,Wei1990} The calculated formation energy of V$_\text{{B}}$ with a clustered (disordered) distribution is predicted to be 0.55 (0.20) eV/atom higher than in uniform distribution. It indicates that V$_\text{{B}}$ tends to uniformly distribute in borophene. This feature can be understood by the occurrence of charge accumulation around six boron atoms neighboring V$_\text{{B}}$, as shown in Figure \ref{chg-B-H}a. The strong Coulomb repulsion between V$_\text{{B}}$ drives them away from one another, and hence a uniform distribution to minimize the total Coulomb interaction. Our discovery of easy V$_\text{{B}}$ formation is in accordance with the recent experimental and theoretical findings of several new phases of two-dimensional borophene, which are probably stabilized by boron vacancies.\cite{Feng2016,Tsafack2016} 

We can also find in Figure \ref{chg-B-H} that in the region far from the defect, the cylinder-like charge density distribution is parallel to the linear B chains (along \emph{x}-direction). In contrast, charge density near the bonds in the zigzag chains is rather low, further confirming that the bond strength of the latter is weaker than the former. The bond lengths of B1-B1 and B2-B2 near V$_\text{{B}}$ are very close to the bulk value 1.65 \AA. Interestingly, B1 and B2 near V$_\text{{B}}$ attract each other and the B1-B2 bond length reduces to 1.66 {\AA} from 1.85 {\AA}. As can been from Figure \ref{chg-B-H}a, this reduction leads to a charge transfer from the dangling bonds of B atoms near V$_\text{{B}}$ to the marginal region near the B-hexagon. Consequently, the bond strength of B1-B2 near V$_\text{{B}}$ is enhanced and the formation energy of V$_\text{{B}}$ is thus lowered in compensation.   

\begin{table*}
\centering
%\begin{ruledtabular}
\caption{\label{tab1} Calculated formation energies $\Delta H$ (eV/atom) of boron vacancy V$_\text{B}$, interstitial \emph{X}$_i$ and substitutional \emph{X}$_\text{B}$ defects (\emph{X}=B, H, C, N and O), bond length $d_\text{B1-B1}$, $d_\text{B2-B2}$,  $d_\text{B1-B2}$, $d_{X-\text{B1}}$ and $d_{X-\text{B2}}$ near defects (\AA). B1 and B2 atoms are colored in green and blue respectively in Figures \ref{chg-B-H} and \ref{chg-N-O}. For the defective systems, the defect concentration is around 3.3\%.}
\begin{tabular}{c|ccccccc}
Systems &  $\Delta H$ (300 K)  & $\Delta H$ (600 K)& $d_\text{B1-B1}$ & $d_\text{B2-B2}$ & $d_\text{B1-B2}$ & $d_{X-\text{B1}}$ & $d_{X-\text{B2}}$ \\
\hline
V$_\text{B}$ & 0.10  & 0.13 &  1.64 & 1.68 & 1.66 & - &-\\
B$_i$ &1.00  & 1.03 & 1.67 & 1.65 & 1.93 &  1.72 & 2.02\\
H$_\text{B}$ & 1.79  & 2.02 & 1.67 & 1.64 & 1.73 & 1.64 & 1.91 \\
H$_{i}$  & -0.44  & -0.21 & 1.73 & 1.61 & 1.99 & 1.20 & 2.84 \\
C$_\text{B}$ & 2.23  & 2.25 & 1.66 & 1.59 & 1.90 & 1.53 & 1.98 \\
C$_{i}$  & 2.85  & 2.87 & 1.69 & 1.66 & 1.89 &1.64 & 1.91 \\
N$_\text{B}$ & 1.54  & 1.87 & 1.75 & 1.62 & 1.86 & 1.48 & 2.29  \\
N$_{i}$  & 0.29  & 0.62 & 1.80 & 1.61 & 1.86 & 1.41 & 2.91 \\
O$_\text{B}$ & -0.79   & -0.46  & 1.98 & 1.60 &1.86 &  1.48 & 2.60\\
O$_{i}$  & -3.02 & -2.68 & 1.77 & 1.59 & 1.82 & 1.42 & 2.77\\
\end{tabular}
%\end{ruledtabular}
\end{table*}

The most stable site for an interstitial B is $t$ site (Figure \ref{chg-B-H}b). The attraction of B$_i$ to its four neighbors leads to a B$_i$-B1 bond length of 1.72 {\AA}, with accumulated charge between them. On the other hand, the B2 atom directly below B$_i$ is pushed downward with a B$_i$-B2 distance of 2.02 {\AA}. The repulsion results in significant local structure distortion around B$_i$, and hence the higher formation energy of B$_i$. 

When H replaces B, it donates partial electrons to its neighboring B atoms. Both B1 and B2 are slightly displaced towards each other by 0.07 {\AA} when compared with their ideal bond length of 1.85 {\AA} (Figure \ref{chg-B-H}c). Since H has only one electron, it causes six unsaturated dangling bonds with its nearest-neighboring B atoms. This is the reason why H$_\text{B}$ is energetically unstable. 
As can be seen from Figure \ref{chg-B-H}d, the most favorable position for interstitial H is $c$ site, the directly below B1 atom moves towards H$_i$ and forms a mixed ionic-covalent bond. By comparison, the two B1 neighbors are displaced slightly away from this B1 atom and yield an equilibrium B1-B1 distance of 1.73 {\AA}. However, the B1-B2 bonds in the hexagonal ring are further strengthened due to the contribution of partial electrons from H$_i$. 

\begin{figure}[H]
\centering
\includegraphics[scale=0.60]{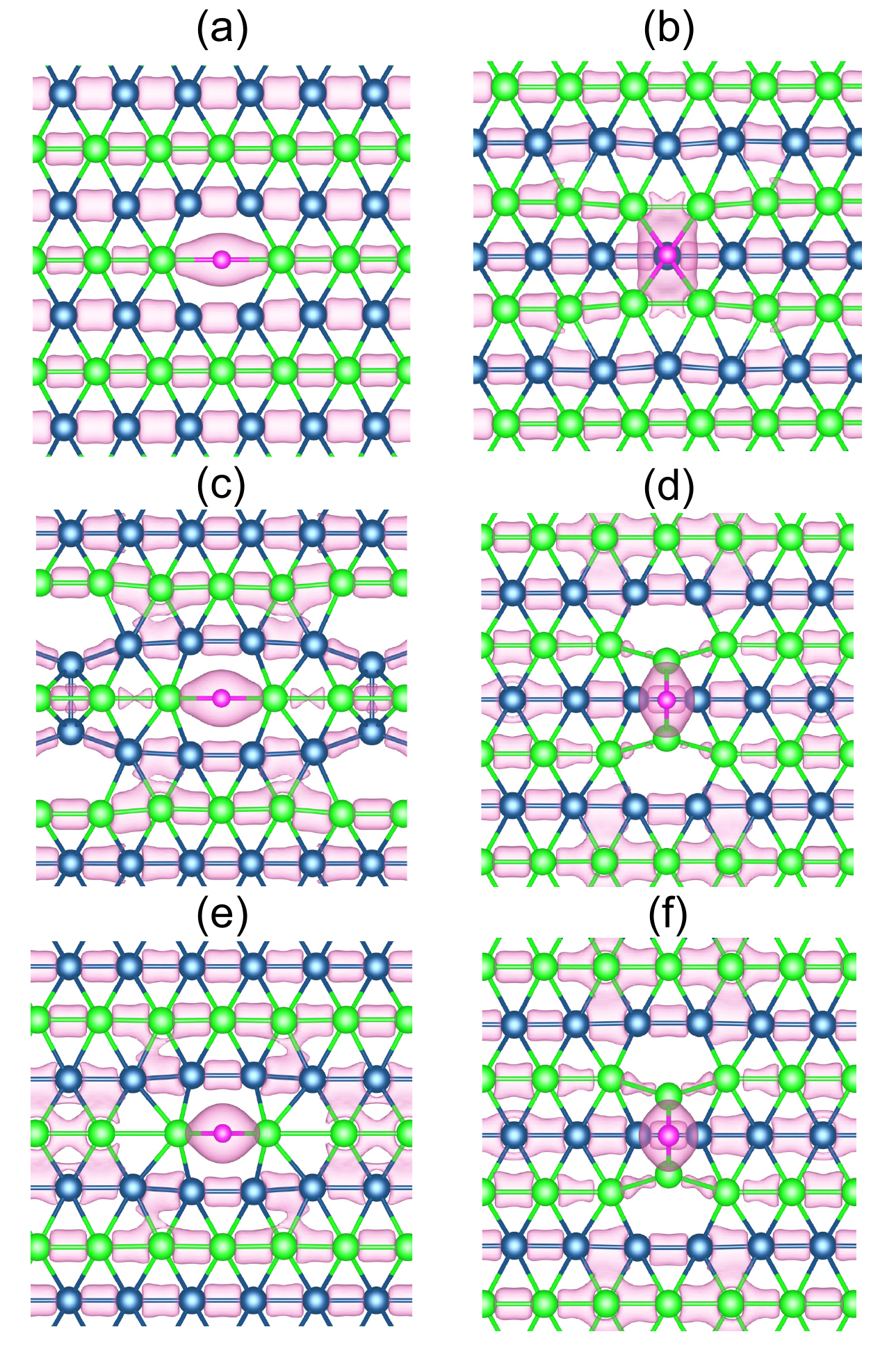}
\caption{\label{chg-N-O}(Color online) Charge density plots of the energetically most stable defective systems consisting of one (a) C$_\text{B}$ and (b) C$_i$, (c) N$_\text{B}$, (d) N$_i$, (e) O$_\text{B}$ and (f) O$_i$ defect, respectively. The B1 and B2 are colored with green and blue balls. The isosurface is 0.12 e/Bohr$^3$. }
\end{figure}

Due to the similarity in covalent radius (0.82 {\AA} for B and 0.77 {\AA} for C), the substitution of B with C atom induces only marginal lattice distortion and charge density redistribution around C$_\text{B}$ atom, as clearly shown in Figure \ref{chg-N-O}a. However, one can find that the calculated formation energy of C$_\text{B}$ is as high as 2.23 eV at 300 K; whereas the value of O$_\text{B}$ is -3.02 eV smaller (Table \ref{tab1}). It should be pointed that the formation energy of $X$-related defect depends on its atomic chemical potential.\cite{Wang2015d} The C atom in the form of graphite has a lower chemical potential (-9.34 eV) than the O atom (-4.93 eV) in oxygen molecule. For purpose of comparison, we further investigate the stability of an isolated C or O atom binding with single B vacancy in borophene respectively. In sharp contrast, the binding energy of C$_\text{B}$ is -5.84 eV, even energetically lower by -1.67 eV than that of O$_\text{B}$. The above explanation still holds true for interstitial C case. The $t$ site is the most stable site for interstitial C. The bond length of C$_i$-B1 is 1.64 {\AA} with typical covalent charge accumulations along these bonds, shorter by around 0.3 {\AA} than that of C$_i$-B2, and thus leading to a four-fold-coordinated bonding configuration (Figure \ref{chg-N-O}b). Furthermore, due to the small value of constant-pressure heat capacity for graphite, the formation energies of both C$_\text{B}$ and C$_i$ are weekly temperature-dependent.

The N$_\text{B}$ defect attracts its two neighboring B1 atoms through strong bonding, resulting in significant local distortion. We see in Figure \ref{chg-N-O}c that a strong reconstruction occur at the B sites next to the N$_\text{B}$. The most favorable interstitial site is $b_y$ site for interstitial N. Compared to the case of N$_\text{B}$, the neighboring B atoms of N$_i$ undergo much smaller offsite displacement, yielding a significantly lower formation energy of N$_i$. With a larger electronegativity than C, both N$_\text{B}$ and N$_i$ capture more electrons from their nearest-neighboring B atoms.

From Figures \ref{chg-N-O}e and \ref{chg-N-O}f, one can find that the features of O-doping are similar in trend to the cases of N-doping. The fact that O has one more valence electron leads to large disparity in formation energies.
One of the most important findings in this study is that the formation energy of O$_i$ is as low as -3.02 eV at room temperature, lower by around 1.0 eV than that of O$_i$ on phosphorene.\cite{Ziletti2015} This means that borophene can be oxidized more easily than phosphorene. More recently, Alvarez-Quiceno \emph{et al}. \cite{Alvarez-Quiceno2017} studied the oxidation mechanism of borophene through first-principles calculations and they found that oxidation of the buckled structure could induce the planar structure with an ordered distribution of vacancies.

The adsorption of H on borophene is also exothermic, adding to the environmental instability of borophene once exposed to air. 
Nevertheless, as can be seen from Table \ref{tab1}, the calculated formation energy of all extrinsic impurities increases monotonously as the temperature increases from 400 K to 800 K, implying that the concentration of these defects can be suppressed in higher-temperature growth. Interestingly, none of the defects considered in our current study produce any low-symmetry Jahn-Teller distortions, as seen in Figures \ref{chg-B-H} and \ref{chg-N-O}. 
Furthermore, from the charge density distribution plotted in Figure \ref{chg-N-O}, one can observe that the ionic bonding character becomes more and more significant when changing from C to O element.

Having evaluated the local structure and stability of various point defects, we now investigate the influence of these defects on the mechanical properties of borophene. We have calculated orientation-dependent Young's moduli $\emph{Y}$ and Poisson's ratio $\nu$. An orientation-dependent strain $\varepsilon$($\theta$) will respond when an uniaxial stress $\sigma$($\theta$) is applied at the angle $\theta$ with respect to the \emph{x}-axis (0$^\circ$ $\leq$ $\theta$ $\leq$ 90$^\circ$). For a 2D crystal, the relationship between the elastic constants and moduli can be given based on the Hooke's law under in-plane stress condition,

\begin{center}
$\left[\begin{array}{c}
\sigma_{xx}\\
\sigma_{yy}\\
\sigma_{xy}
\end{array}\right]=\left[\begin{array}{ccc}
C_{11} & C_{12} & 0\\
C_{12} & C_{22} & 0\\
0 & 0 & C_{66}
\end{array}\right]\left[\begin{array}{c}
\varepsilon_{xx}\\
\varepsilon_{yy}\\
2\varepsilon_{xy}
\end{array}\right]$,
\par\end{center}
where \emph{C}$_{ij}$ (\emph{i},\emph{j}=1,2,6) is the in-plane stiffness tensor and is equal to the second partial derivative of strain energy $E_{S}$ with respect to strain $\varepsilon$. Using the standard Voigt notation,\cite{Andrew2012} \emph{i.e.}, 1-\emph{xx}, 2-\emph{yy}, and 6-\emph{xy}, it can be written as $C=(1/S_{0})(\partial^{2}E_{S}/\partial\varepsilon_{i}\partial\varepsilon_{j})$, where $S_{0}$ is the equilibrium area of the system. Generally, in first-principles calculations, the \emph{C}$_{ij}$ can be obtained based on the following formula,\cite{Zhang2015}
%\begin{center}
\begin{equation}\label{eq4}
E_{s}=\frac{1}{2}C_{11}\varepsilon_{xx}^{2}+\frac{1}{2}C_{22}\varepsilon_{yy}^{2}+C_{12}\varepsilon_{xx}\varepsilon_{yy}+2C_{66}\varepsilon_{xy}^{2},
\end{equation}
where the tensile strain is defined as $\varepsilon=\frac{a-a_{0}}{a}$, \emph{a} and \emph{a}$_0$ are the lattice constants of the strained and strain-free structures, respectively. Applying uniaxial strain $\varepsilon$ applied along \emph{x} (\emph{y}) direction leads to $\varepsilon_{yy}$=0 ($\varepsilon_{xx}$=0) and $E_{s}=1/2C_{11}\varepsilon_{xx}^{2}$ ($1/2C_{22}\varepsilon_{yy}^{2}$). It may be worth mentioning here that the dimension was allowed to shrink in the orthogonal direction. Then, the relevant elastic constant $C_{11}$ and $C_{22}$ can be acquired from the coefficient of the quadratic term by fitting the data of elastic strain energy $E_s(\varepsilon)$ as a function of strain $\varepsilon$ using a quadratic polynomial. Finally, we can obtain $E_{s}=(1/2C_{11}+1/2C_{22}+C_{12})\varepsilon_{xx}^{2}$ when equi-biaxial strain is applied. In order to calculate the elastic stiffness constants, the $E_{s}$ as a function of $\varepsilon$ in the strain range -2\% $\leq$ $\varepsilon$ $\leq$ 2\% with an increment of 0.5\% are investigated. The elastic constants \emph{C}$_{ij}$ can be obtained by post-processing the VASP calculated data using the VASPKIT code.\cite{vaspkit}

\begin{figure}[H]
\centering
\includegraphics[scale=0.5]{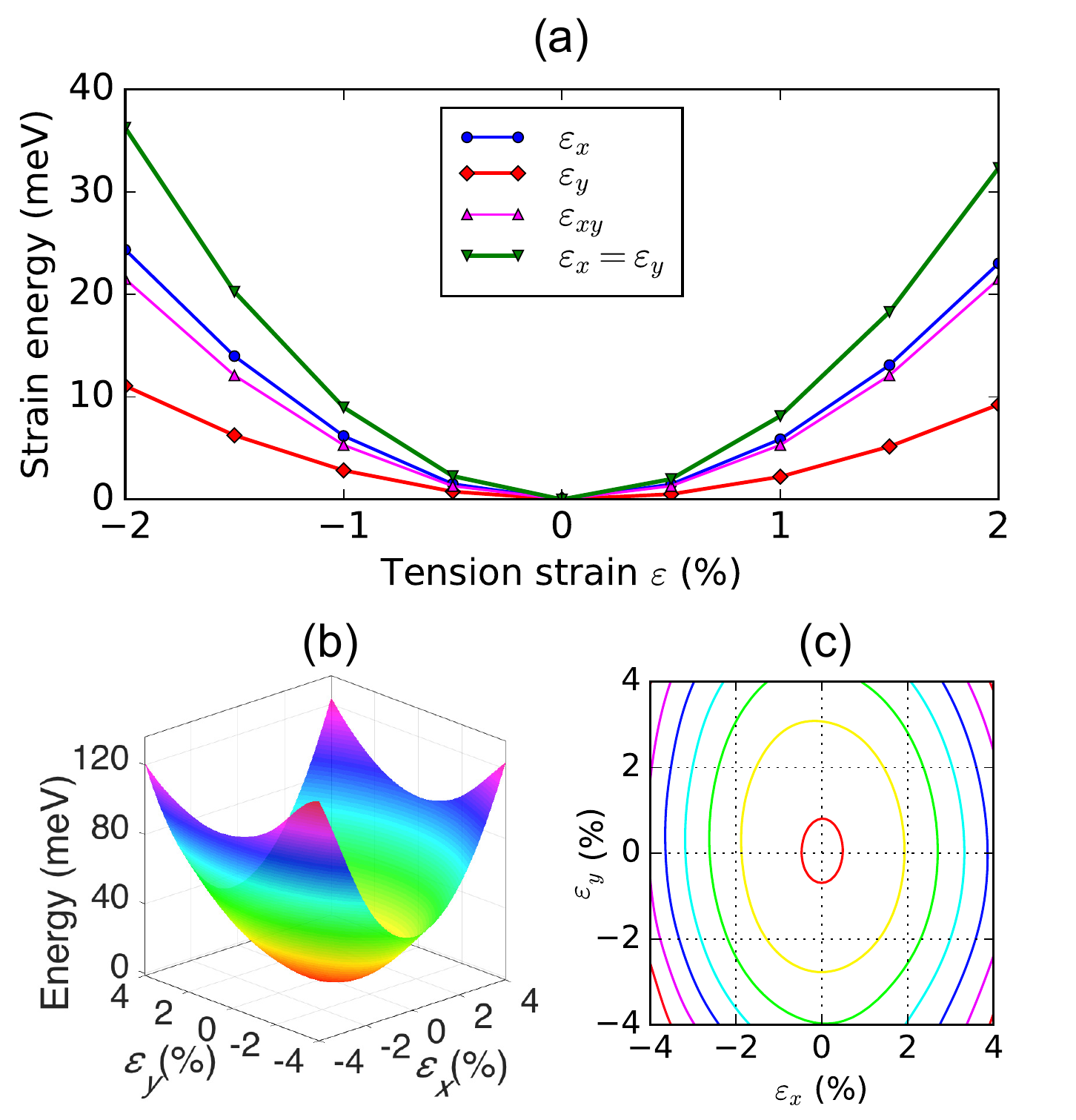}
\caption{\label{tension}(Color online) (a) Changes in the strain energy $E_{s}$ per unit cell as a function of strain $\varepsilon$ for freestanding borophene sheet. (b) Three-dimensional energy surface and (c) contour plots on the mesh of data points (\emph{a}$_x$,\emph{a}$_y$) used for the strain energy calculations.}
\end{figure}

For the purpose of illustration, we take pristine borophene as an example and present its strain-energy versus strain-curve in Figure \ref{tension}.  
For the orthogonal 2D system, the strain parallel to $\theta$ direction $\varepsilon_{\parallel}$, and the strain perpendicular to $\theta$ direction $\varepsilon_{\perp}$ induced by the unit stress \bm{$\sigma$}($\theta$) (\bm{$\left|\sigma\right|$}=1) are expressed as,\cite{Yakobson2015}

\begin{eqnarray}
&&\varepsilon_{\parallel}=\frac{C_{11}s^{4}+C_{22}c^{4}-2C_{12}c^{2} s^{2}}{C_{11}C_{22}-C_{12}^2}+\frac{c^{2} s^{2}}{C_{66}},
\end{eqnarray}
and
\begin{eqnarray}
&&\varepsilon_{\perp}=\frac{-C_{12}(s^{4}+c^{4})+(C_{11}+C_{22})c^{2} s^{2}}{C_{11}C_{22}-C_{12}^{2}}-\frac{c^{2} s^{2}}{C_{66}},
\end{eqnarray}

respectively, where $s=sin(\theta)$ and  $c=cos(\theta)$.

Then the expressions for the orientation-dependent Young's modulus $\emph{Y}$($\theta$) and Poisson's ratio $\nu$($\theta$) are derived as
\begin{eqnarray}
&&Y(\theta)=\frac{\sigma}{\varepsilon_{\parallel}} \nonumber \\
&&=\frac{C_{11}C_{22}-C_{12}^{2}}{C_{11}s^{4}+C_{22}c^{4}+(\frac{C_{11}C_{22}-C_{12}^{2}}{C_{66}}-2C_{12})c^{2}s^{2}},
\end{eqnarray}
and 
\begin{eqnarray}
&&\nu=-\frac{\varepsilon_{\perp}}{\varepsilon_{\parallel}} \nonumber \\
&&=\frac{C_{12}(c^{4}+s^{4})-(C_{11}+C_{22}-\frac{C_{11}C_{22}-C_{12}^{2}}{C_{66}})c^{2}s^{2}}{C_{11}s^{4}+C_{22}c^{4}+(\frac{C_{11}C_{22}-C_{12}^{2}}{C_{66}}-2C_{12})c^{2}s^{2}},
\end{eqnarray}
respectively. For a 2D crystal, its shear modulus is $G$=$C_{66}$.

\begin{figure}[H]
\centering
\includegraphics[scale=0.22]{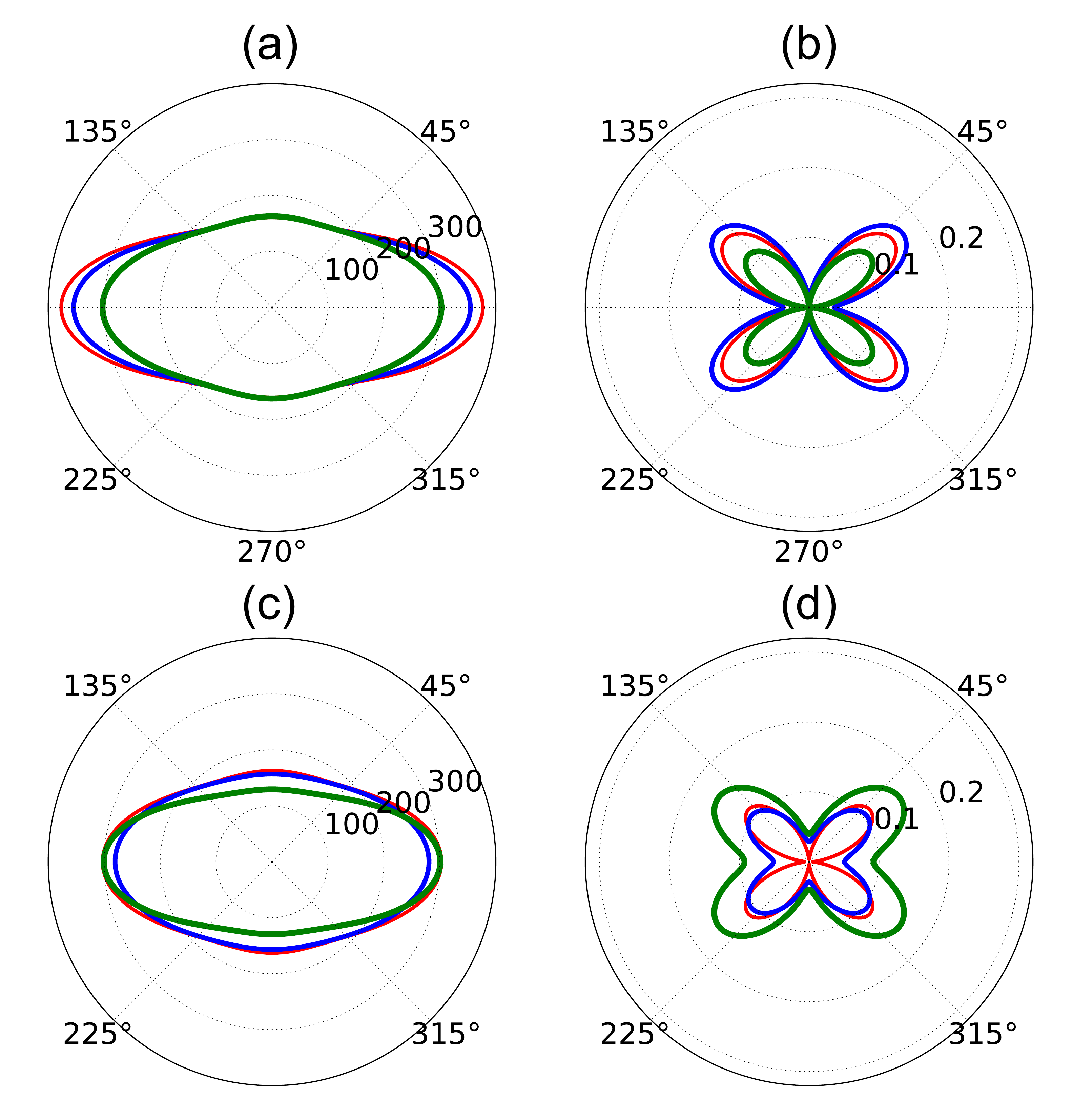}
\caption{\label{mech-host}(Color online) The calculated orientation-dependent Youngs's modulus $Y(\theta)$ [(a) and (c)], Poisson's ratio $\nu(\theta)$ [(b) and (d)]. The red, blue and green lines in (a) and (b) represent the systems without defect, with one V$_\text{B}$, and with one B$_{i}$ respectively, and those in (c) and (d) represent the systems with uniform, disordered and clustering V$_\text{B}$ distribution respectively. For the defective systems, the defect concentration is around 3.3\%.}
\end{figure}

As can be seen from Table \ref{tab2} and Figure \ref{mech-host}a, the calculated orientation-dependent Young's modulus $\emph{Y}$($\theta$) of borophene decreases monotonically from the maximum value of 377 GPa$\cdot$nm along the $x$ direction ($\theta$=0$^\circ$) to the minimum value of 161 GPa$\cdot$nm  along the $y$ direction ($\theta$=90$^\circ$), in agreement with previous theoretical values of 398 and 170 GPa$\cdot$nm.\cite{Mannix2015} The highly anisotropic nature of elastic properties in borophene can be understood from the fact that the B-B bond along $x$ has the highest strength than any other directions. Interestingly, the Posson's ratio $\nu$($\theta$) first increases to its maximum value of 0.154 at $\theta$$\approx$39$^\circ$ from a value of 0.005 at $\theta$=0$^\circ$, and then decreases to a minimum value of 0.002 at $\theta$=90$^\circ$. In contrast, Mannix \emph{et al.} reported that borophene structure has an intrinsic negative Poisson's ratio of -0.04 (-0.02) in the $x$ ($y$) direction by using PBE approach without van der Waals dispersion correction.\cite{Mannix2015} As pointed out in Sec. II, the incorporation of the non-bonding van der Waals interaction into density
functional theory calculations is necessary in order to provide more accurate description of the geometrical properties of two-dimensional materials. As a matter of fact, our PBE calculations reproduce the negative Posson's ratio character for borophene (Table \ref{tab2}).

We see in Table \ref{tab2} all the point defects considered will decrease the maximum Young's modulus (\emph{i.e.}, along \emph{x} direction). However, their influence on Posson's ratio is more complicated. We define the reduction rate of Young's modulus to evaluate the relative changes in Young's modulus of the defective system, $\delta=1-\frac{Y_{M}(X)}{Y_{M}(host)}$, where $Y_{M}(X)$ and $Y_{M}(host)$ are the maximum value of Young's modulus of the configurations with and without defect. We classify the defective systems into three groups based on the magnitude of $\delta$.

\begin{figure}[H]
\centering
\includegraphics[scale=0.24]{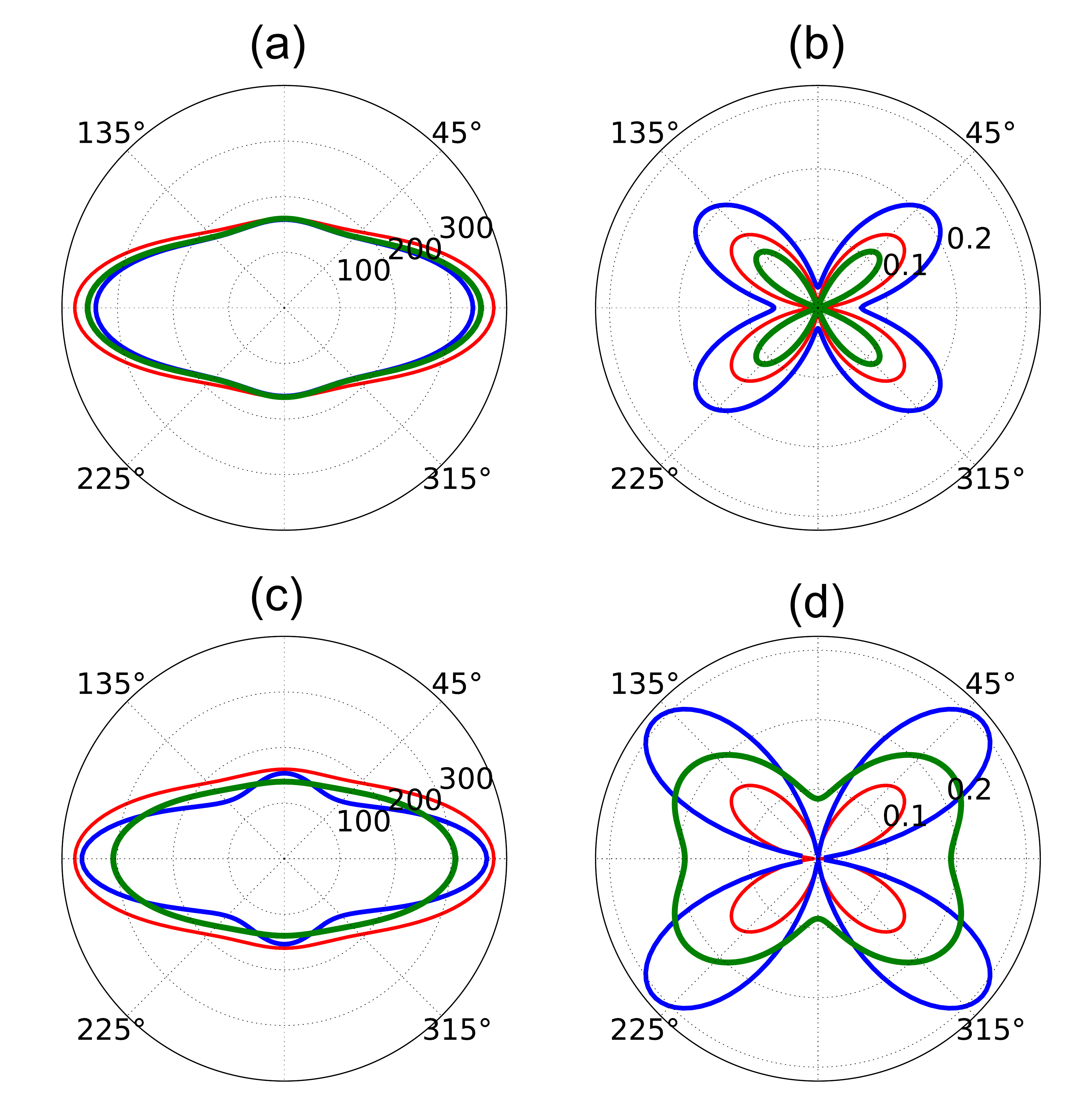}
\caption{\label{mech-H_C}(Color online) The calculated orientation-dependent Youngs's modulus $Y(\theta)$ [(a) and (c)], Poisson's ratio $\nu(\theta)$ [(b) and (d)]. The red, blue and green lines in (a) and (b) represent the systems without defect, with one H$_\text{B}$, and with one H$_{i}$ respectively, and those in (c) and (d) represent the systems without defect, with one C$_\text{B}$, and with one C$_{i}$ respectively. For the defective systems, the defect concentration is around 3.3\%.}
\end{figure}

\emph{A. $\delta$ < 10\% for the systems consisting of one B$_i$, H$_i$, H$_\text{B}$ or C$_\text{B}$ defect}. The small values of $\delta$ could be attributed to the common feature that all of these defects induce small local structure distortions. However, the change in the Poisson's ratio are very different. H$_\text{B}$ or C$_\text{B}$ greatly enhance the Poisson's ratio of borophene. H$_i$ and C$_\text{B}$ even lead to negative Poisson's ratio values of -0.136 and -0.063 along $x$ direction. Our results suggest that negative Poisson's ratio in borophene could be realized or strengthened by doping selected impurities wisely.   

\begin{figure}[H]
\centering
\includegraphics[scale=0.25]{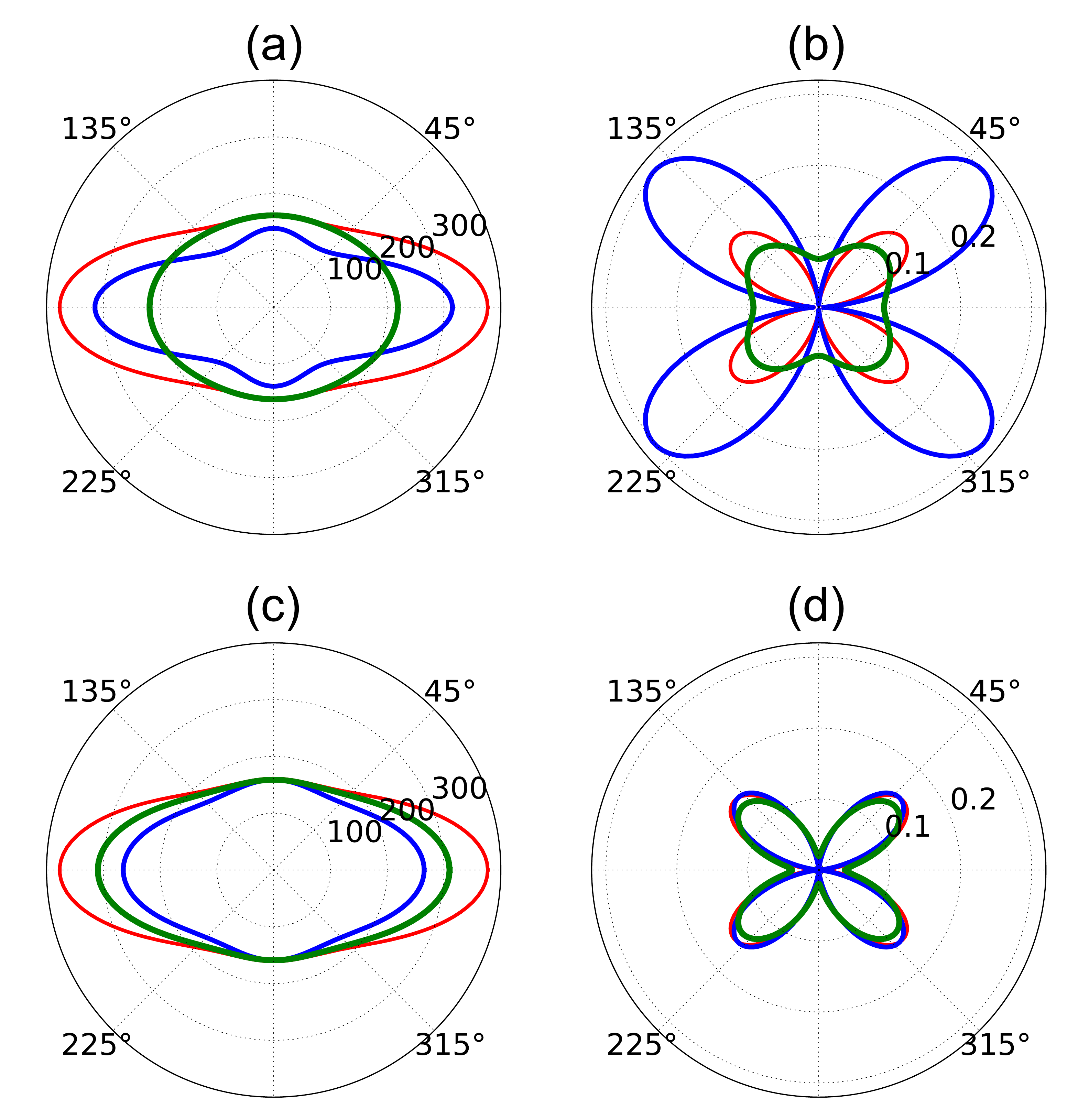}
\caption{\label{mech-N_O}(Color online) The calculated orientation-dependent Youngs's modulus $Y(\theta)$ [(a) and (c)], Poisson's ratio $\nu(\theta)$ [(b) and (d)]. The red, blue and green lines in (a) and (b) represent the systems without defect, with one N$_\text{B}$, and with one N$_{i}$ respectively, and those in (c) and (d) represent the systems without defect, with one O$_\text{B}$, and with one O$_{i}$ respectively. For the defective systems, the defect concentration is around 3.3\%.}
\end{figure}

\begin{table*}
\centering
%\begin{ruledtabular}
\caption{\label{tab2} Calculated elastic stiffness constants C$_{ij}$ (GPa$\cdot$nm), Youngs's modulus $Y$ (GPa$\cdot$nm) and Poisson's ratio $\nu$ of defective borophene along \emph{x} and \emph{y} directions. The maximum values of Poisson's ratio $\nu_{M}$ are also presented. All defects are supposed to have a uniform distribution unless otherwise stated. V$_\text{B}^\text{d}$ and V$_\text{B}^\text{c}$ represent disorder and clustering distributions of B vacancies. For the defective systems, the defect concentration is around 3.3\%.}
\begin{tabular}{c|ccccccccc}
Systems   & C$_{11}$ & C$_{22}$ & C$_{12}$ & C$_{66}$  & \emph{Y}$_x$ & \emph{Y}$_y$ & $\nu_{x}$ & $\nu_{y}$ &$\nu_{M}$  \\
\hline
host    & 377 & 161 &1 &84  & 377  &162 & 0.005 &0.002 & 0.154\\
host$^a$       & 405 & 172 & -1 & 96  &405 & 172& -0.006 &-0.003 & 0.117\\
host\cite{Mannix2015} & 398 & 170 & -7 & 94  &398 & 170& -0.040 &-0.020 & -\\
%host\cite{Wang2016}  & 379 & 162 & -2 & 87  &379 & 162& -0.012 &-0.005 & -\\
V$_\text{B}$   & 303 &  163 & 1 & 85 & 304  &163 & 0.009 &0.005 &0.114\\
V$_\text{B}^\text{d}$  & 281 & 157 &  8 & 85 & 281  &156 & 0.049 &0.027 & 0.106\\
V$_\text{B}^\text{c}$   & 302 & 130 & 12 &71 & 301  &130 & -0.096 &-0.042 & 0.159\\
B$_i$   & 355 & 163 &8 & 82   & 356  &163 & 0.035 &0.016 & 0.170\\
H$_\text{B}$  & 340 & 159 & 10 & 74  & 339  &158 & 0.062 &0.029 & 0.215\\
H$_{i}$   & 357 & 162 & -22 & 81   & 354  &160 & -0.136 &0.062 &0.116 \\
C$_\text{B}$ & 365 & 154 & -10 & 56   & 365  &153 & -0.063 &-0.026 & 0.308\\
C$_{i}$  & 313 & 141 & 27 & 71   & 308  &138 & 0.195 &0.088  & 0.228\\
N$_\text{B}$   & 315 & 139 & 1 & 53  & 316  &140 & 0.004 &0.002 & 0.302 \\
N$_{i}$    & 220 & 163 & 15 & 80  & 219  &162 & 0.091 &0.067 & 0.118\\
O$_\text{B}$  & 265 & 159 & -1 & 73  & 264  &159 & -0.006 &-0.003 & 0.153\\
O$_{i}$  & 310 & 159 & 6 & 82  & 310  &159 & 0.040 &0.020 &0.140\\
\end{tabular}
\leftline{$^a$ PBE-calculated values without van der Waals dispersion correction. }
%\end{ruledtabular}
\end{table*}

\emph{B. 10\%<$\delta$<20\% for C$_i$, O$_i$, N$_\text{B}$ and V$_\text{B}$}. There are more significant local distortions near these defects. Similar to H$_i$ and C$_\text{B}$, both C$_i$ and N$_\text{B}$ increase the anisotropy of Poisson's ratio in borophene. 
It is noteworthy that besides the species and types of defects, the distribution of defects also affects the mechanical properties of borophene. We take B vacancy as an example and calculate the elastic properties of three configurations with uniform, clustered and disordered distribution of V$_\text{B}$, respectively. The calculated results are presented in Table \ref{tab2} and Figures \ref{mech-host}c and \ref{mech-host}d. The Young's modulus of borophene is decreased by around 6 \% along \emph{x} direction with a disordered V$_\text{B}$ distribution; while with a clustered V$_\text{B}$ distribution it is lowered by 20 \%  compared with the system in which V$_\text{B}$ are uniformly distributed. As expected, recent theoretical studies predict that the concentration of V$_\text{B}$ has an important effect on the mechanical anisotropy properties of borophene.\cite{Wang2016,Zhang2017}

\emph{C. $\delta$>30\% in x direction for N$_i$ and O$_\text{B}$ defects}. Interestingly, these two types of defects have negligible effects on the Young's modulus along $y$ direction. Since both N$_i$ and O$_\text{B}$ have low formation energies (seen Table \ref{tab1}), especially the latter one, they are expected to degrade mechanical properties of borophene more significantly.   

\section{Grain boundaries}

\begin{figure}[H]
\centering
\includegraphics[scale=0.4]{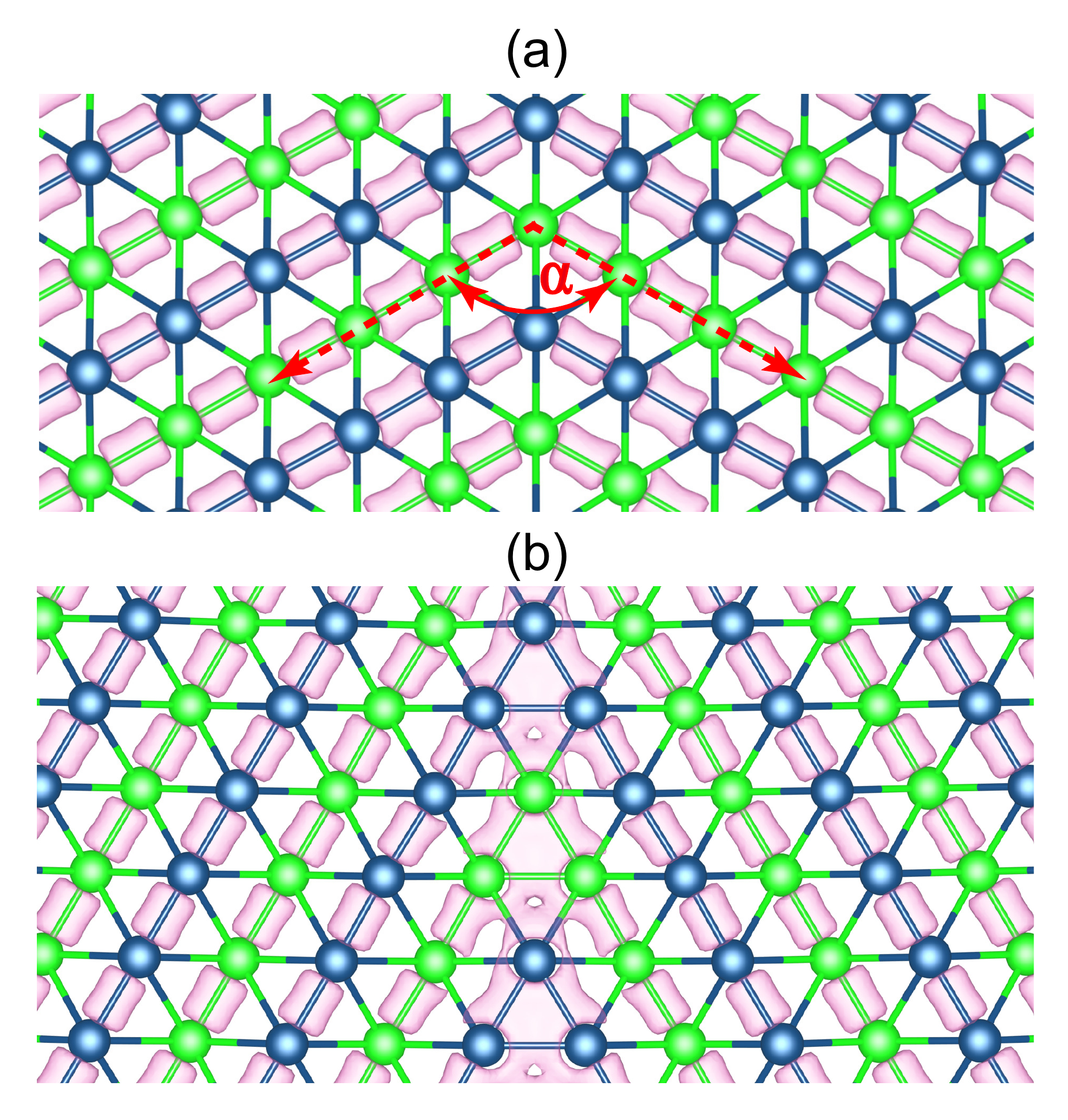}
\caption{\label{gb}(Color online) The $\Sigma$5 tilt grain boundaries in borophene with electron density isosurface of 0.12 eV/bohr$^3$, (a) GB1 and (b) GB2.}
\end{figure}

In Figures \ref{gb}a and \ref{gb}b we display two symmetric tilt $\Sigma$5 grain boundaries in borophene and denote them as GB1 and GB2 respectively. As illustrated in Figure \ref{struct}, the GB1 and GB2 structures can be made by applying a mirror-reflection operation on the unit cell shadowed in blue and red colors. The axis of symmetry are \emph{k} and \emph{m}, the angle $\theta$ of these axis relative to the \emph{x}-axis are 120$^\circ$ and 30$^\circ$ respectively. Their GB tilt angles $\alpha$ are 120$^\circ$ and 60$^\circ$ respectively. To describe the stability of GB structure, we define the GB formation energy as 

\begin{equation}\label{eq_gb}
\Delta E_G=\frac{E_{tot}(GB)-E_{tot}(host)}{2L},
\end{equation}
where $E_{tot}(X)$ and $E_{tot}(host)$ represent the total energies of a GB supercell and a supercell of perfect lattice with the same size.
$L$ is the length of the supercell along the grain boundary.

The calculated GB formation energies are 0.16 eV/{\AA} for GB1 and 0.06 eV/{\AA} for GB2. In Figure \ref{gb}, one can see that the valence charge densities accumulate in the region of GB2 while there is no significant change accumulation near the GB1. As a result, the B1-B2 bond strength near GB2 is stronger than that near GB1. For this reason, it is expected that the GB2 dominates over GB1 in borophene. To determine the role of GBs on the mechanical properties of borophene, we calculated the Youngs's modulus parallel ($Y_{\parallel}$) and perpendicular ($Y_{\perp}$) to the GB direction. For the GB1 structure, 
$Y_{\perp}$=220 ($\theta$=30$^\circ$) GPa$\cdot$nm and $Y_{\parallel}$=168 ($\theta$=120$^\circ$) GPa$\cdot$nm; while for GB2, $Y_{\perp}$=166 ($\theta$=120$^\circ$) GPa$\cdot$nm and $Y_{\parallel}$=204 ($\theta$=30$^\circ$) GPa$\cdot$nm.
It should be noted that the GB1 and GB2 directions are perpendicular to each other. Compared to pristine borophene, the Youngs's modulus decreases by 9.5 \% across $k$ axis for GB1, and by 16.1 \% across $m$ axis for GB2. 
Then, an important question arise: why negligible changes of Youngs's modulus are observed along particular directions in these two GB models, both of which are parallel to the $k$ axis? A closer look at Figure \ref{gb} tells that the B1-B2-B1-$\cdot$$\cdot$$\cdot$ like zigzag chain along the $k$ axle is parallel to these two directions. The changes in the structure of zigzag chain are extremely small upon the introduction of GB1 or GB2. Therefore, both GBs have little effect on the mechanical properties along or cross the GB directions. 

\section{Summary}
In conclusion, we have investigated the stability of lattice defects
as well as their influence on the mechanical anisotropy of borophene  using density functional first-principles calculations including van der Waals correction. We find that the formation energies of point defects, B vacancy, interstitial H, interstitial N, substantial O and interstitial O are less than 0.3 eV at room temperature. In addition, 
we have also examined two high angle tilt grain boundary structures whose formation energies are as low as 0.16 eV/{\AA} and 0.06 eV/{\AA}. These point defects can derease severely the in-plane Young's moduli of borophene at high concentraion. Our results demonstrate borophene has high chemical activity and is not stable in the air, and this drawback has to be overcome before its real applications. It is also found that the substitution of C or N for B can significantly increase the Poisson's ratio and the adsorbed H or substitutional B can bring about negative Poisson's ratio in borophene. The two grain boundaries, on the other hand, exert only limited influence. 

\section{Acknowledgement}
We thank Y.C. Liu for valuable discussions and acknowledge the financial support of the Special Scientific Research Program of the Education Bureau of Shaanxi Province, China (Grant No. 15JK1531), and the National Natural Science Foundation of China (Grant Nos. 11304245 and 61308006).
\nocite{*}

\bibliographystyle{achemso}
%\bibliography{references}
\providecommand{\latin}[1]{#1}
\providecommand*\mcitethebibliography{\thebibliography}
\csname @ifundefined\endcsname{endmcitethebibliography}
  {\let\endmcitethebibliography\endthebibliography}{}

\end{document}